# On the Uncertainty in X–Ray Cluster Mass Estimates

Christophe Balland[1]

University of California

Center for Particle Astrophysics

301 LeConte Hall

Berkeley, CA 94720, USA

and

Alain Blanchard[2]

Observatoire Astronomique de Strasbourg

11 Rue de l'Université

67 000 Strasbourg, France

---

[1]Also Observatoire Astronomique de Strasbourg and Université d'Aix-Marseille II, France

[2]Also Université Louis Pasteur de Strasbourg, France

# ABSTRACT

We study the uncertainty in galaxy cluster mass estimates derived from X-ray data assuming hydrostatic equilibrium for the intra cluster gas. Using a Monte-Carlo procedure we generate a general class of mass models allowing very massive clusters. We then compute the corresponding temperature profiles through the equation of hydrostatic equilibrium and compare them to observational data on some clusters. We find several massive clusters that pass the observational constraints, with integrated masses varying in a quite wide range. The resulting accuracy of the mass estimates is rather poor, larger than what is generally claimed. Despite the fact that the mass profile can mathematically be determined exactly from the knowledge of the temperature and surface brightness profiles, we find that very accurate measurements of both quantities are required to determine the actual mass with moderate accuracy. We argue that the tight constraints on cluster masses previously obtained come from the fact that a too restricted class of mass density profiles has been investigated so far, without serious physical motivations. Applying our procedure to the Perseus and then to the Coma cluster, we find that the improvement of the observational constraints results in a quite modest improvement in the accuracy of the mass estimate. For Coma, using the best current available data, we end up with a factor two of uncertainty in the mass within the Abell radius. This uncertainty rapidly increases at further radius.

*Subject headings:* galaxies: clustering — galaxies: intergalactic medium — galaxies: X-rays

## 1. Introduction

The determination of the mass of galaxy clusters is of major importance in cosmology. Dynamical estimates of cluster mass provide one of the few evidences that the mean density of the universe is at least 10% of the critical density, in excess of what is detected in individual galaxies. The discovery of the X–ray emitting gas in cluster cores has revealed the presence of a large baryonic component. Under the assumption of hydrostatic equilibrium (HE), this gas offers a way to determine the radial mass distribution of clusters. This procedure yields in principle much more accurate results than the determination based on optical data (Sarazin 1986). The method has been widely used in the past and claimed accurate mass estimates have been inferred from such analyses (e.g., Cowie, Henriksen & Mushotzky 1987; Hughes, Gorenstein & Fabricant 1988; Hughes 1989 – hereafter H89; Eyles et al. 1991; Gerbal et al. 1992; Watt et al. 1992; Durret et al. 1994; Elbaz, Arnaud & Böhringer 1995). H89 used a combination of available data, both optical and X–rays, to derive simultaneously the amount of mass in the gas component and in the dark matter in the Coma cluster. These quantities have been refined by Briel, Henry & Böhringer (1992) – BHB hereafter – by use of recent ROSAT observations. Similar analyses have been performed on other clusters: A2256 (Henry, Briel & Nulsen 1993; Miyaji et al. 1993), A2163 (Elbaz et al. 1995), A426 (Cowie et al. 1987; Eyles et al. 1991) among others. Although these mass estimates have reinforced the evidence for the presence of dark matter, it has been recently outlined that the fraction of baryonic material in clusters seems to be inconsistent with the overall baryon density of the universe predicted by the standard theory of cosmic nucleosynthesis in an Einstein-de Sitter ($\Omega_0 = 1$) model (D. White & Fabian 1995; S. White et al. 1993; D. White et al. 1994). The case of Coma has been the subject of much attention as baryons make up a substantial fraction of the total mass ($\approx 8h^{-3/2}\%$; S. White et al. 1993), but in some other cases, the baryonic fraction appears to be similar or even higher (Eyles et al. 1991; Durret et al. 1994; D. White et al. 1994; Elbaz et al. 1995).

In this Letter we investigate whether accurate mass estimates can be obtained from the HE equation (note that our aim is not to discuss the relevance of the assumption of HE in cluster central regions). Actually, the possible uncertainty on mass estimates is generally believed to be small (Loewenstein 1994; Schindler 1995; Evrard, Metzler & Navarro 1995), as soon as the temperature is reasonably well known. We consider a new approach to address this question, based on the use of a Monte-Carlo procedure. We expose here our main results and conclusions. The details of our calculations as well as the impact of the current uncertainty in mass estimates on the problem of the baryon fraction in the universe will appear in a forthcoming paper (Balland & Blanchard 1995).

## 2. Reliability of Mass Estimates

Under the hydrostatic assumption, the temperature of the X–ray gas traces the gravitational potential of the whole cluster. This may be written as (assuming spherical symetry):

$$T(r) = \frac{1}{\gamma_g + \gamma_T} \frac{GM(r)}{r}. \tag{1}$$

In principle the slope of the radial gas distribution $\gamma_g$ and the temperature gradient $\gamma_T$ are accessible through the observations of $T(r)$ and of the luminosity profile $s(r)$ due to the thermal bremsstrahlung emission of the gas, provided that the observed quantities are deprojected. The emissivity profile of clusters, in the absence of cooling flows, is well fitted by a standard $\beta$-profile (Cavaliere & Fusco-Femiano 1976):

$$s(r) \propto [1 + (r/r_c)^2]^{-3\beta+1/2}, \tag{2}$$

where $r_c$ denotes the gas core radius and $\beta$ is a fitting parameter (physically the ratio of the galaxy kinetic energy to the gas energy). Once deprojected, $s(r)$ gives access to the radial distribution of the gas with satisfying accuracy. One might therefore expect from equation (1) that the mass at a given radius is uncertain to a precision comparable to the uncertainty on the temperature. One should however be extremely cautious when trying to use equation (1) to infer $M(r)$ from the oberved temperature profile $T(r)$, as equation (1) is actually a differential equation. It happens that the solutions obtained through the HE equation are actually very sensitive to the boundary condition (as noted by some authors, see, e.g., Loewenstein 1994). Several rules have been used in the past to limit the solutions that are actually investigated. Some authors have argued that solutions for which the temperature diverges or reaches zero at some finite radius have to be ruled out, i.e. only critical solutions for which the pressure goes smoothly down to zero at infinity have to be kept (e.g., Loewenstein 1994). We would like to emphasize that these assumptions severely limit the class of solutions under investigation without being well motivated on a physical basis: the hot gas in clusters is probably heated by shocks during the phase of non-linear collapse and a shock front is expected. The assumption of hydrostatic equilibrium is certainly wrong for radii larger than the shock radius which may be of the order of the virial radius or less. Clearly, a reliable boundary condition would require a delicate assumption on the discontinuity at the front shock.

Besides, in most analyses to date, only one physical scale is introduced in the mass profile. The obvious first physical scale entering the problem is the core radius of the gas component. Actually, some authors have noted that the mass may be more concentrated in the cores of clusters than in the outer parts (Gerbal et al. 1992; Loewenstein 1994). This tendency has been confirmed by the mass determination within cluster centers through gravitational lensing effect (Hammer 1991; Mellier, Fort & Kneib 1993). Therefore the mass profile might differ considerably from that of the gas component and might possess several physical scales. For instance, from the observational point of view, although it is known that the temperature is rather constant up to a few core radii (see, e.g., the Coma cluster; Hughes et al. 1988), it is natural to expect that the gas temperature drops to zero far away from the center in the case of an isolated bounded cluster. If the temperature does not present a strong discontinuity, but rather decreases smoothly after some radius this may introduce another scale length in the problem. It is thus important to allow for more freedom when one investigates mass estimates of X–ray clusters.

## 3. Method

We have thus investigated the possibility that the mass contains several scales, having in mind that the mass density profile may become *shallower* in the outer parts of clusters. We therefore assume a simple mass distribution for which the density profile in the central region of clusters is given by the profile corresponding to the isothermal case (the gas distibution following in this case an isothermal $\beta$-profile):

$$\rho(r) \propto \frac{3 + (r/r_c)^2}{[1 + (r/r_c)^2]^2} \qquad (3.1)$$

and the central dark matter density $\rho_0$ is left as a free parameter. In the outer parts of the cluster, beyond some transition radius $r_t$ we assume that the mass density is a power law with index $\alpha$, which is also a free parameter:

$$\rho(r > r_t) \propto (r/r_c)^\alpha. \qquad (3.2)$$

$\alpha$ is allowed to vary between $-3$ and $0$ (the asymptotic slope of the isothermal solution beeing $-2$). We do not attempt to argue that such slopes are realistic from a physical point of view: although the slope of the dark matter distribution is known to depend on the power spectrum as well as on the density of the universe (Crone, Evrard & Richstone 1994), some of the profiles we use are rather extreme and may appear unrealistic. We fully appreciate this point but as we investigate the possible uncertainties from a purely phenomenological point of view, we do not want to add any theoretical prejudice to our analysis if it can not be directly tested by the observations. We also investigate profiles that are peaked in the central region (referred later as "centrally peaked profiles"). We add in this case a term $1/[1+(r/r_c)^2]^n$ to equation (3.1), where $n$ is taken between 2.25 and 3. We therefore generate mass models according to (3) by a Monte-Carlo procedure, computing the corresponding temperature profiles through equation (1) and comparing them to various sets of available data. We keep any solutions satisfying the observational constraints without imposing any further restrictions based on some (uncertain) physical assumptions. The observational data used are the emissivity profile and the distance up to which the emission has been confidently detected as well as the information on the temperature profile. This procedure is efficient as long as the number of constraints is not very high nor of extremely good quality, otherwise the required number of models to investigate is prohibitive. In the following, the quantities chosen at random are: 1) the total mass $M_A$ within the Abell radius which is allowed to vary between half and six times the mass corresponding to the isothermal solution within the same radius, 2) the central dark matter density $\rho_0$ varying within at most 10% of the isothermal value and 3) the slope $\alpha$. The trial is kept only if a radius $r_t$ can be chosen consistently with equation (3).

## 4. Results

Integrating equation (1) with fixed central temperature as boundary condition, we find that the isothermal solution is obtained for a very precise value of the central binding density $\rho_0$. Any departure from this peculiar fine-tuned value leads to a dramatic behaviour of the solution. A difference as small as 0.5% yields temperature profiles that fall to zero within a few core radii or that diverge rapidly. This extreme sensitivity of the hydrostatic equation to the boundary conditions while noticed has not been emphasized and its consequences have not been fully appreciated. In addition the temperature solutions exhibit a surprising result: the temperature decreases in the outer part of the most massive profiles rather than increasing as one would expect from a naive examination of the hydrostatic equation (1).

We present the results of our simulations in the particular case of two well-known clusters: A426 (Perseus) and A1656 (Coma) .

### 4.1. Perseus

In figure 1, we give examples of emission-weighted projected temperature profiles fitting roughly the temperature data from SPACELAB-2 (Eyles et al. 1991; solid bars). Also shown are new data from ASCA (Arnaud et al. 1994; dashed bars). As our aim is to compare with previous mass estimates (e.g., Loewenstein 1994), we do not try to fit our models to this new set of data. The core radius and the $\beta$ parameter have been taken from Eyles et al. (1991) in good agreement with Jones & Forman (1984) results: $r_c \sim 9.1'$ and $\beta \sim 0.57$. Note that Perseus has a central cooling flow that we do not try to model. We find that the mass within $1.3 h_{50}^{-1}$ Mpc may vary between $\sim 4.1 \times 10^{14} h_{50}^{-1} M_\odot$ and $\sim 1.3 \times 10^{15} h_{50}^{-1} M_\odot$, i.e. a factor 3 of uncertainty (throughout this work, we assume $h_{50} = 1$). This has to be compared with the range $4.4 - 4.8 \times 10^{14} h_{50}^{-1} M_\odot$ derived by Loewenstein (1994), who integrates the hydrostatic equation inward from boundary condition taken at infinity. At the Abell radius ($3 h_{50}^{-1}$ Mpc), where no data are available, the mass appears to be almost unconstrained (more than a factor 10 of uncertainty). If we now impose that the temperature is non-zero within $70'$ – the radius up to which significant X-ray emission is detected (Eyles et al. 1991) – this reduces to a factor of 1.5 the uncertainty of the mass at 1.3 Mpc and to a factor 3 at the Abell radius. Therefore the uncertainty remains quite substantial. It may be thought that this is due to the fact that data are of rather poor quality, and that an improvement will result in a substantial reduction of the uncertainty. As we will illustrate this in the case of Coma, for which the data are significantly better, the actual improvement is rather modest.

## 4.2. Coma

For Coma, we first apply the constraints used by H89: 1) that the temperature does not drop to zero within $\sim 50'$ from the cluster center and 2) that the averaged synthetic temperature within the EXOSAT and TENMA beams ($0.75^o$ and $3^o$ respectively) yields the observed values ($8.5 \pm 0.5$ KeV and $7.5 \pm 0.2$ KeV – at the $2\sigma$ level – respectively). For each of our model temperature profiles passing criterion 1), we calculate the average temperature weighted on the (projected) luminosity profile in order to reproduce exactly the observational procedure. Because the temperature is allowed to be zero after only $50'$, we find a large uncertainty in the mass at the Abell radius (a factor 3) and more than one order of magnitude at $5h_{50}^{-1}$ Mpc (for comparison, H89 found only a factor 3 of uncertainty in the mass within $5h_{50}^{-1}$ Mpc with his models – see table 5 in H89; S. White et al. 1993 quoted a 20% uncertainty at the Abell radius). Moreover, the masses we derive within these radii are typically larger than those inferred by H89. This is a direct consequence of using shallower profiles. However, ROSAT detected gas emission of Coma up to $\sim 95'$ with an emissivity well fitted by equation (2) with $\beta \sim 0.75$ and $r_c \sim 10.5'$ (BHB). We thus impose now that the temperature does not drop to zero within $95'$ and we apply the same procedure as before. In general "centrally peaked profiles" are preferred. Some of the synthetic profiles passing the new constraints are shown in figure 2. The corresponding mass profiles appear in figure 3 (solid lines) as well as some mass profiles allowed when the constraint that the temperature is non-zero within $95'$ is relaxed to $50'$ (dashed lines). Despite the fact that the emission has been detected by ROSAT twice as far as before, the spread in the mass at the Abell radius is still a factor $\sim 2$: M($< 3h_{50}^{-1}$Mpc) $\sim 1.2 - 2.5 \times 10^{15} M_\odot$, and reaches up to a factor $\sim 7$ at 5 Mpc: M($< 5h_{50}^{-1}$Mpc) $\sim 1.3 - 9.2 \times 10^{15} M_\odot$. Very few mass models from our Monte-Carlo samples actually pass the observational constraints because of the accuracy of the measurements. Models which satisfy the average temperature in EXOSAT and TENMA beams at the $3\sigma$ level are much more common, and the mass dispersion is larger, of the order of 3 at the Abell radius. Finally, we use the most constraining temperature data available for Coma, i.e., GINGA data (Jones & Forman 1992). Note that these data are not fully consistent with the constraints from TENMA and EXOSAT. We present in figure 4 emission-weighted projected profiles from our sample fitting these data (solid lines). Once the underlying mass profile is determined in our simulation, the two parameters relevant to the fit are $\rho_0$ and the central temperature $T_0$. The dotted line corresponds to a modified "isothermal" profile, i.e. an isothermal profile with a value of $\rho_0$ slightly different from the isothermal value. The mass associated with this profile within the Abell radius is $1.95 \times 10^{15} h_{50}^{-1} M_\odot$. The maximum mass we obtain within the Abell radius is $3 \times 10^{15} h_{50}^{-1} M_\odot$ (solid lines on figure 4.). Also shown is the model temperature used by Makino (1993) (dashed line), who derives a mass of $1.6 \times 10^{15} h_{50}^{-1} M_\odot$ within the same radius using the same information on the luminosity profile as we do, namely the ROSAT emissivity profile for Coma. (Note that the lowest mass we get at the Abell radius by our Monte-Carlo procedure is $1.8 \times 10^{15} h_{50}^{-1} M_\odot$). Taking these results together, we find a spread in the allowed mass of about a factor 2 within the Abell radius. It is clear that even with improved data, the uncertainty in the mass remains large at this radius, of the order of a factor 2. It is important to keep in mind that this is minimum value, as more complex profiles might allow a wider mass range.

## 5. Conclusion

We have generated a large number of mass models and derived the corresponding temperature profiles via the equation of hydrostatic equilibrium. In our analysis, a general class of models has been investigated and we have applied this procedure to the specific cases of the Coma and Perseus clusters, though our main conclusions are valid for any cluster. We find from available observations of the temperature and X–ray emission that the mass of clusters is much worsely constrained by the hydrostatic method than what is usually be-lieved: on our low mass range we recover the typical masses that were previously inferred, but also find more massive profiles that fit identically well the observations. Even with the most constraining observations we find that the uncertainty on the total mass remains of the order of a factor 2 within the radius up to which data are available. We believe that this number is a lower limit and the uncertainty could be larger for other clusters for which observations of poorer quality are available. We conclude that cluster masses cannot be determined accuratly by the sole use of the equation of hydrostatic equilibrium even when X-ray data of good quality are available. Accordingly, cluster masses may have been underestimated so far, which, if true, would alleviate the problem of the baryonic fraction in clusters.

We wish to thank Joe Silk, Gary Mamon, Yoel Rephaeli, David Valls-Gabaud and Jim Bartlett for numerous fruitful discussions regarding the results presented in this paper. CB acknowledges financial support from the Center for Particle Astrophysics of the University of California at Berkeley.

---

This manuscript was prepared with the AAS LaTeX macros v4.0.

Figures captions

Fig. 1.— Emission-weighted projected temperature profiles for the Perseus cluster. Data are from Eyles et al. (1991) (solid) and new ASCA measurements from Arnaud et al. (1994) are shown for comparison (dashed)

Fig. 2.— Synthetic temperature profiles for Coma. All the profiles pass the constraint imposed by the measurements of TENMA and EXOSAT satellites and are non-zero within $95'$ from the cluster center

Fig. 3.— Mass profiles corresponding to the temperature profiles of figure 2 (solid lines). The dashed lines correspond to temperature profiles allowed to drop to zero beyond $50'$ from the cluster center and satisfying the constraints imposed by TENMA and EXOSAT (as in Hughes 1989). Also shown is the range of allowed mass obtained by Briel et al. (1992) at $5h_{50}^{-1}$ Mpc from a different set of mass density profiles

Fig. 4.— Emission-weighted projected temperature profiles fitting GINGA data (solid lines). The dotted line is a modified isothermal profile (see text) and the dashed line is the model of Makino (1993). The mass within the Abell radius in our models is higher by a factor $\sim 2$ than the mass inferred within the same radius by Makino

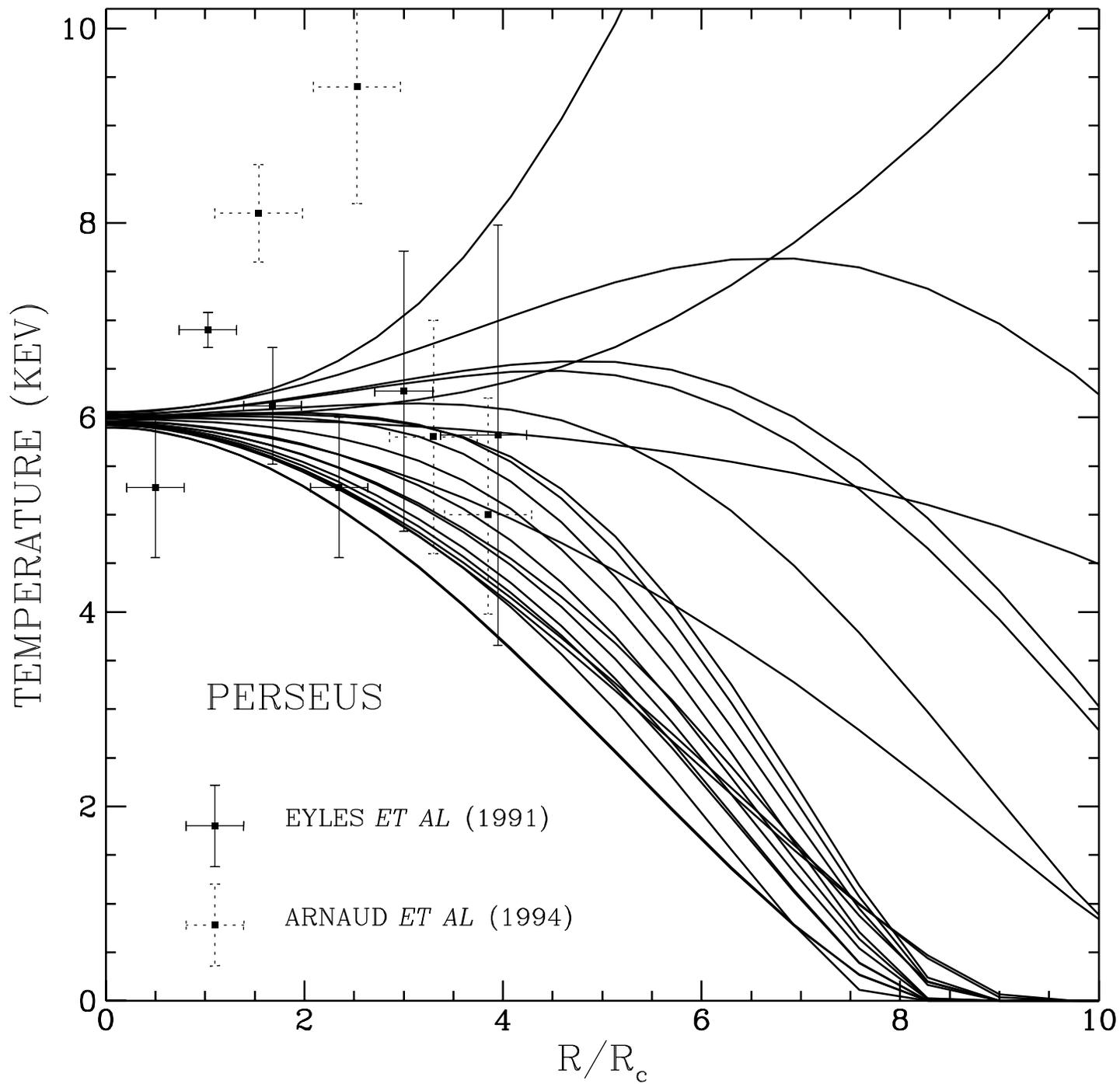

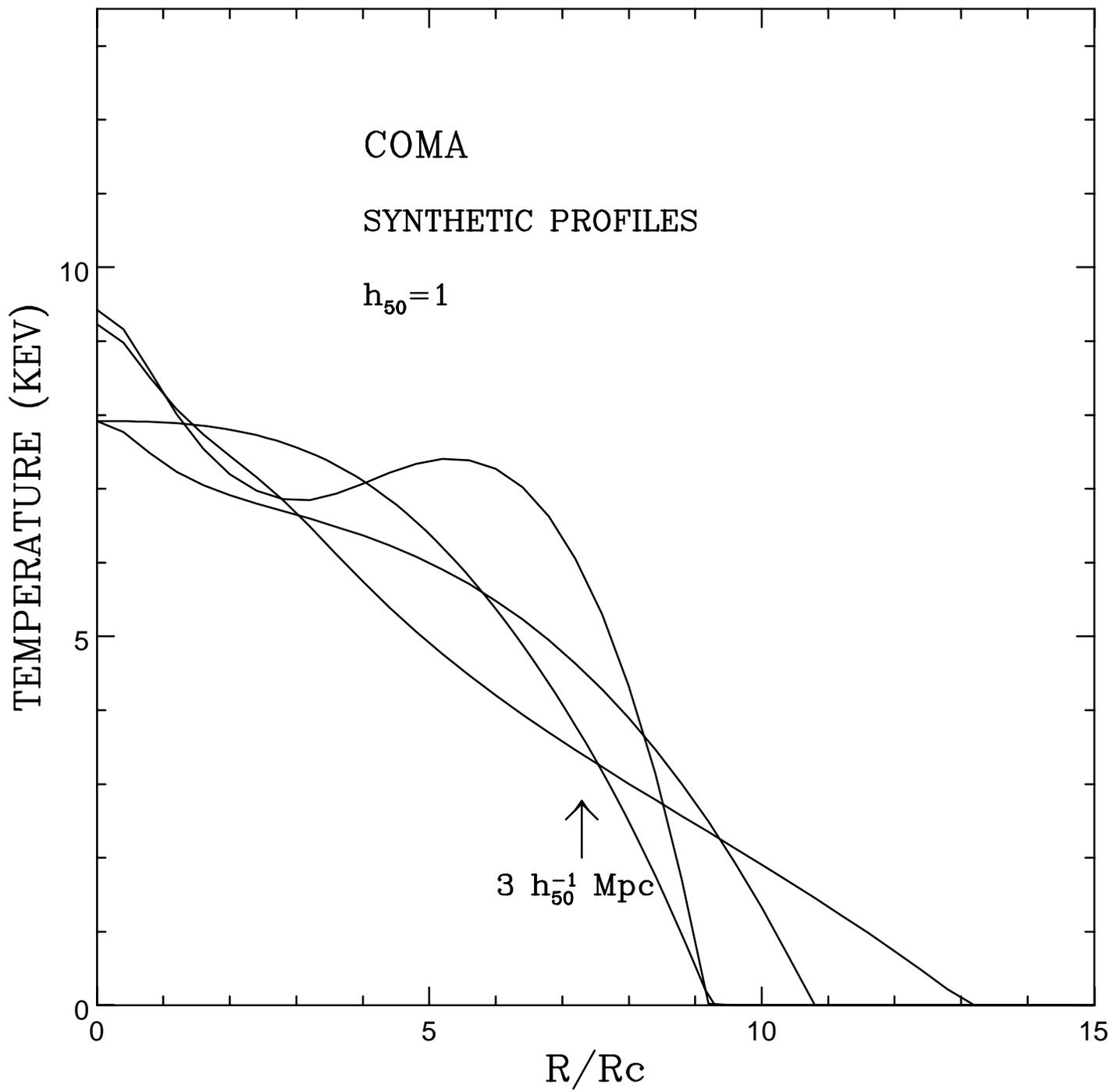

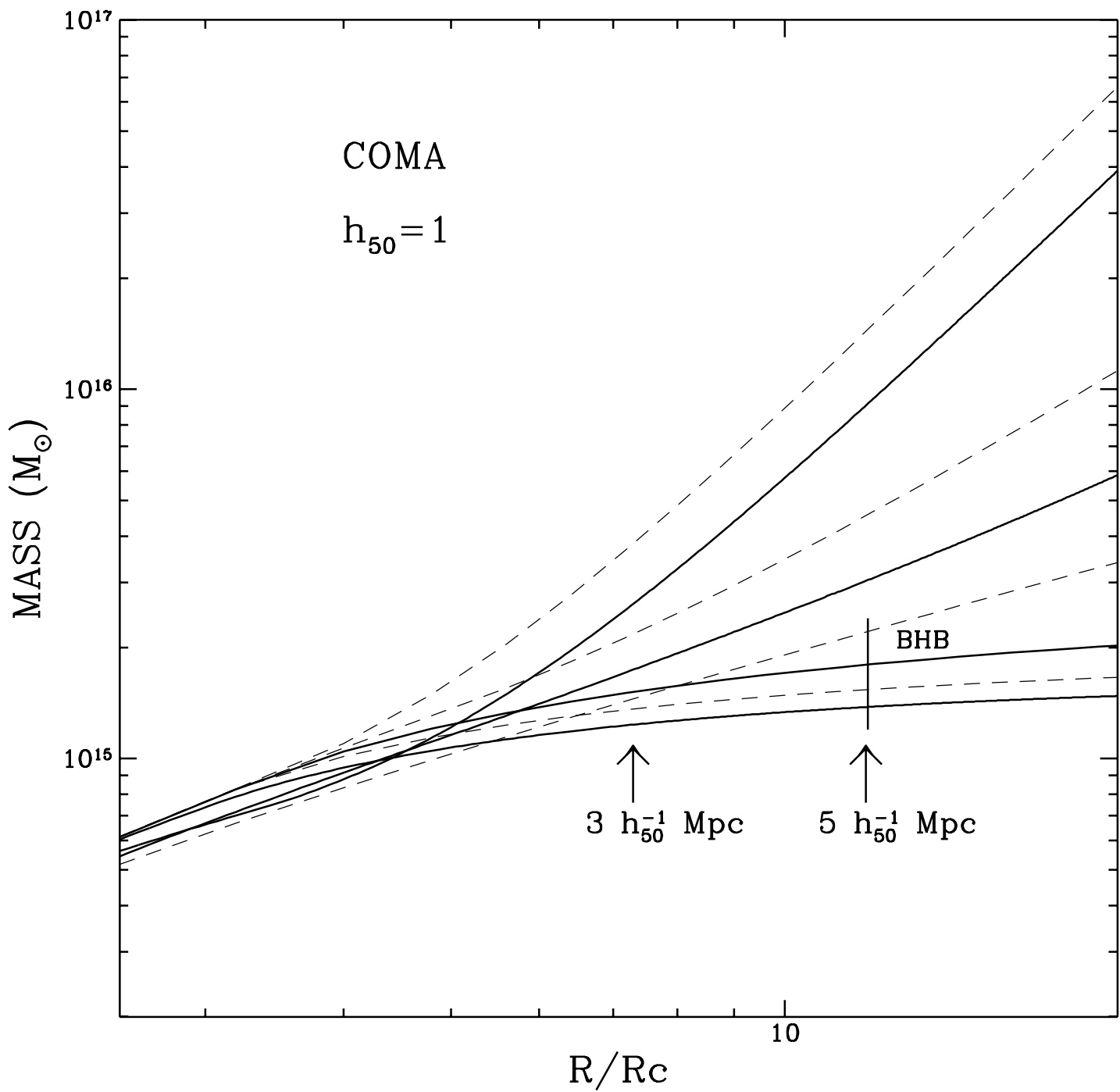

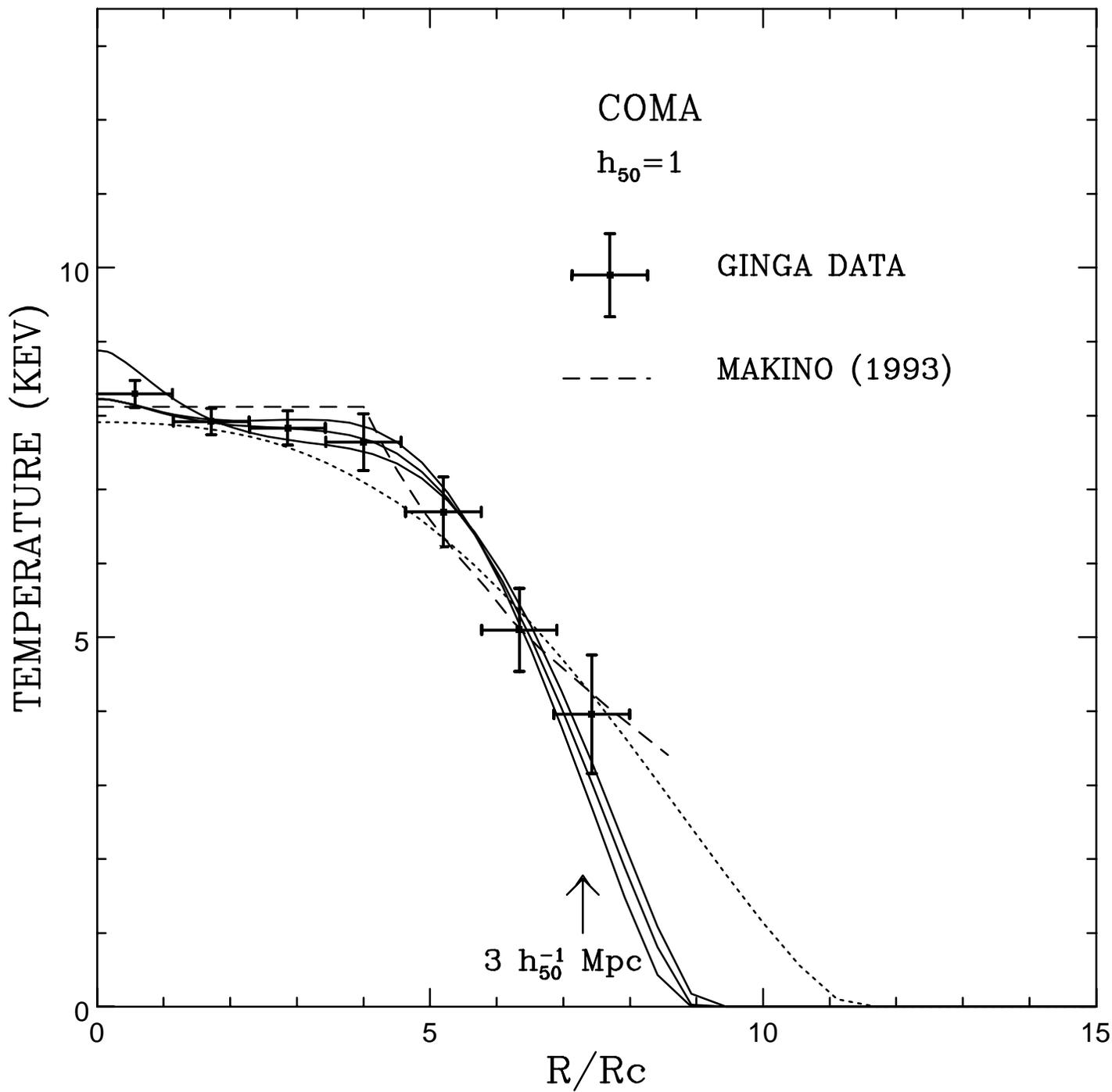